\documentclass[sn-mathphys]{sn-jnl}
\usepackage[utf8]{inputenc}
\usepackage{graphicx}
\usepackage{xr}
\usepackage{ulem}
\usepackage{textcomp}
\usepackage{gensymb}
\hypersetup{colorlinks,
citecolor=black,
filecolor=black,
linkcolor=black,
urlcolor=black
}

\begin{document}

\title[Automatic Parameter Selection for Electron Pytchography via BO]{Automatic Parameter Selection for Electron Ptychography via Bayesian Optimization}
\author[1]{\fnm{Michael C.} \sur{Cao}}
\author[2]{\fnm{Zhen} \sur{Chen}}
\author*[3]{\fnm{Yi} \sur{Jiang}}
\author*[1]{\fnm{Yimo} \sur{Han}}
\affil[1]{Department of Materials Science and NanoEngineering, Rice University, Houston, TX, USA 77005}
\affil[2]{School of Materials Science and Engineering, Tsinghua University, Beijing 100084, China}
\affil[3]{Advanced Photon Source, Argonne National Laboratory, Lemont, IL, USA 60439}

\abstract{
Electron ptychography provides new opportunities to resolve atomic structures with deep sub-angstrom spatial resolution and studying electron-beam sensitive materials with high dose efficiency. In practice, obtaining accurate ptychography images requires simultaneously optimizing multiple parameters that are often selected based on trial-and-error, resulting in low-throughput experiments and preventing wider adoption. Here, we develop an automatic parameter selection framework to circumvent this problem using Bayesian optimization with Gaussian processes. With minimal prior knowledge, the workflow efficiently produces ptychographic reconstructions that are superior than the ones processed by experienced experts. The method also facilitates better experimental designs by exploring optimized experimental parameters from simulated data.}
    
\maketitle

\section{Introduction}
Ptychography is a computational imaging method that has gained great interests in the electron microscopy community\cite{yang2016simultaneous,gao2017electron,jiang2018electron,chen2021electron}. The technique was first proposed by Hoppe in 1969\cite{hoppe1969beugung} and re-invigorated in recent years with the the developments of fast electron detectors\cite{ophus2014recording,tate2016high,mir2017characterisation} that can rapidly collect tens of thousands of diffraction patterns in less than a minute. Various iterative reconstruction algorithms have been developed to retrieve the scattering potentials of the sample and the wave function of the illumination from intensity measurements\cite{thibault2008high,guizar2008phase,maiden2009improved}. It has been demonstrated that electron ptychography can break the Abbe diffraction limit of imaging systems\cite{maiden2011superresolution} and set a new world record in spatial resolution (0.39 Å) in atomically thin two-dimensional (2D) materials\cite{jiang2018electron}. As one of the phase-contrast imaging techniques, electron ptychography also has high dose-efficiency for low-dose imaging ranging from low-dimensional nanomaterials\cite{chen2020mixed,song2019atomic} to biological specimens\cite{pelz2017low,zhou2020low}. An even more critical breakthrough is that electron ptychography can inversely solve the long-standing problem of multiple scattering in thick ($>$20 nm) samples and enables a lattice-vibration-limited resolution (0.2 Å)\cite{chen2021electron}, as well as three-dimensional depth sectioning\cite{chen2021electron, chen2021three}. 

Despite its great success in achieving record-breaking resolution, ptychography remains a niche technique in electron microscopy due to many practical challenges in both experimental setup and data analysis. In particular, there exist many types of parameters that significantly influence image quality and need to be carefully selected for different data or applications. Because modern ptychography recovers object structures via solving an iterative inverse problem, the recovered information strongly depends on the  \textit{reconstruction parameters}, which are related to physical modeling of the data as well as optimization algorithms. In addition,  \textit{experimental parameters} such as scan step size and probe defocus also limits the best image quality of a given measurement. Due to the virtually infinite possibilities and complex trade-offs between various parameters, it is practically impossible to design and optimize ptychography experiments by searching the entire parameter space. In most literature\cite{jiang2018electron,song2019atomic,chen2020mixed,chen2021electron}, both reconstruction and experimental parameters are manually selected based on scientists’ experiences about the sample or the instrument. This not only introduces biases to scientific conclusions that are drawn from the reconstruction but greatly reduces the throughput of the overall workflows, resulting in a high barrier for general researchers to adopt the technique. 

Here we present a general framework for fully automatic parameter tuning in electron ptychography by leveraging Bayesian optimization (BO) with Gaussian processes\cite{rasmussen2003gaussian} -- a popular strategy for global optimization of unknown functions. Using experimental ptychography data and state-of-the-art reconstruction algorithms, we demonstrated that our approach can automatically  produce high-resolution images after exploring only 1\% of the entire reconstruction parameter space. We also  optimized experimental parameters for ultra-low electron dose levels, providing insights for more robust experimental designs that further to enhance ptychography's usability. Instead of relying on human intuition and judgment, automatic parameter selection promotes objective and reproducible protocols, paving the way for fully autonomous experiments and data processing for ptychography applications.

\section{Results}
\subsection{Bayesian Optimization with Gaussian Process}
Bayesian optimization with Gaussian process is frequently used to find global maxima and minima of a black-box function that is unknown and expensive to evaluate. The technique has been used in a wide variety of applications in machine learning\cite{bergstra2011algorithms,brochu2010tutorial}, Monte Carlo simulation\cite{mahendran2012adaptive}, and autonomous controls in microscopy experiments\cite{duris2020bayesian,zhang2021aberration,roccapriore2021physics}. In general, BO consists of three steps: 1) compute a surrogate function that models the true objective function based on sampled points, 2) determine the next point(s) to be sampled based on an acquisition function, 3) evaluate the objective function at the corresponding points. The surrogate function is described by kernel functions, which affect the periodicity, smoothness, and length scales of the objective function. Maximizing the acquisition function provides the points that automatically balance the trade-off between finding the extrema (exploitation) or reducing the uncertainty in the surrogate (exploration). 

\begin{figure}
    \centering
    \includegraphics[width = 1.0\linewidth]{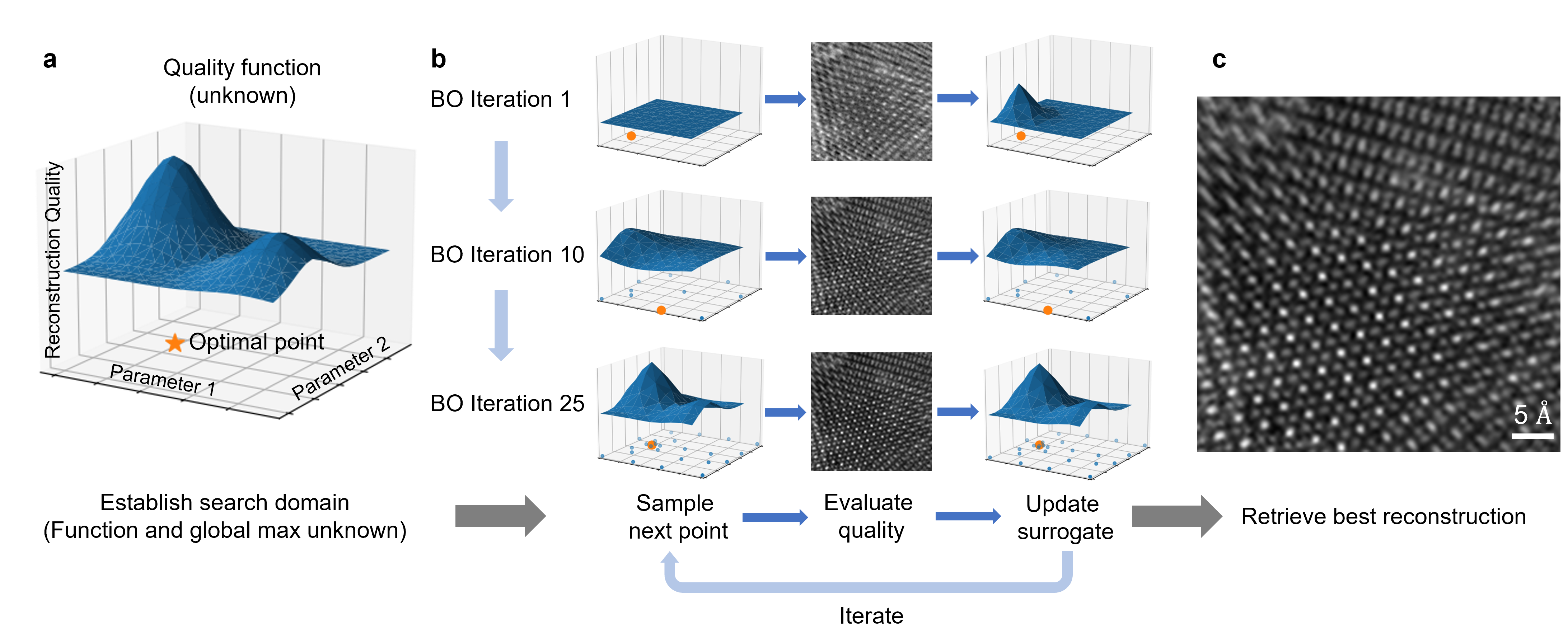}
    \caption{Schematic of automatic reconstruction tuning with Bayesian optimization. (a) The process aims to find the best ptychographic reconstruction by optimizing an unknown quality function that is data-dependent in general. (b) Bayesian optimization loop strategically determines the next point (indicated in orange) to sample, performs ptychographic reconstruction, and then updates the surrogate model based on the image quality. As the number of iterations increases, the surrogate model becomes closer to the true quality function and more points around the optimum are exploited. (c) The image with the best quality during BO is retrieved as the final reconstruction.}
    \label{fig:schematic}
\end{figure}

In ptychography, we utilize BO to optimize an objective function that evaluates reconstruction quality and varies for each dataset. Figure \ref{fig:schematic} illustrates the complete workflow. The initial set of ptychography reconstructions are generated based on randomly chosen parameters. Based on these reconstructions, Gaussian process models a surrogate function, and candidate points thereafter are chosen according to the acquisition function. The following reconstructions are performed with these parameters, the quality is measured, and the surrogate and acquisition functions are updated. The updated acquisition function then suggests the next candidate set and the process is iterated (Figure \ref{fig:schematic}b). After sufficient iterations, the set of parameters that generate the highest quality reconstruction is determined (Figure \ref{fig:schematic}c).

\subsection{Automatic Reconstruction Parameter Tuning}
To demonstrate BO as an efficient framework for automatic selection of reconstruction parameters, we apply the approach on an experimental dataset of bilayer MoSe$_2$/WS$_2$ which is publicly available from ref. \cite{chen2020mixed}. Ptychographic reconstructions were carried out using the least square maximum likelihood (LSQ-ML) algorithm\cite{odstrvcil2018iterative} implemented in the PtychoShelves package\cite{wakonig2020ptychoshelves}, which incorporates many advanced techniques such as mixed-states ptychography\cite{thibault2013reconstructing}, position correction\cite{odstrvcil2018iterative}, variable probe correction\cite{odstrcil2016ptychographic}, batch update\cite{odstrvcil2018iterative}, and multislice ptychography\cite{tsai2016x, chen2021electron}. These features play crucial roles in previous works that successfully achieved dose-efficient and large field of view (FOV) imaging\cite{chen2020mixed,jiang2021achieving} as well as the deep sub-angstrom spatial resolution of thick crystalline materials\cite{chen2021electron}. Because different parameters have different computational costs per iteration, we perform time-limited reconstructions that terminate after reaching a time threshold specified by the user. This provides more practical comparisons by balancing the trade-offs of parameters such as the number of probe modes and batch size. Assuming strong phase approximation, which generally works well for 2D materials, the complete parameter space consists of eight discrete parameters and a total number of 4800 possible combinations. Detailed descriptions of each reconstruction parameter are provided in the Methods section and supplementary Table 1.

\begin{figure}
    \centering
    \includegraphics[width = 1.0\linewidth]{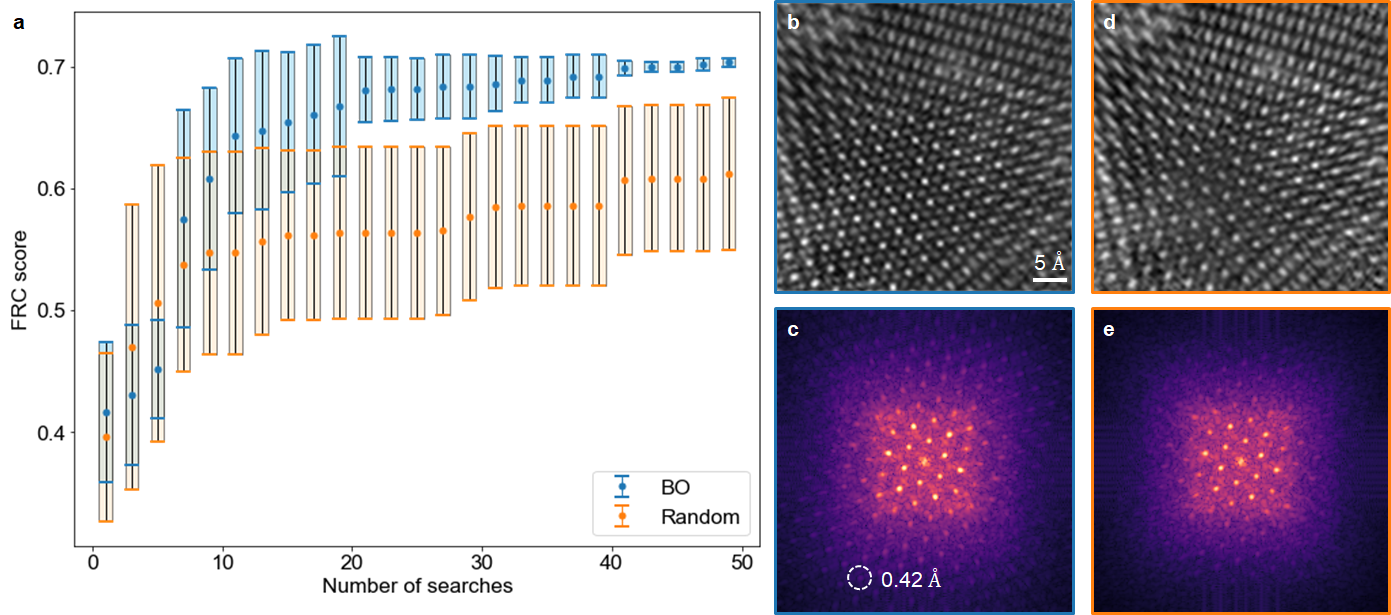}
    \caption{Performance of Bayesian optimization versus a random search. (a) Plot of FRC score behavior over number of parameter searches using BO versus a random search strategy. In 10 trials, BO consistently outperforms random search with higher FRC score and lower uncertainty as the number of parameter searches increases. (b,d) Best reconstruction for one trial after 50 parameter searches using (b) Bayesian optimization and (d) random search. (c,e) Diffractograms of reconstructions in (b) and (d), respectively.}
    \label{fig:bo vs random}
\end{figure}

\begin{figure}
    \centering
    \includegraphics[width = 1\linewidth]{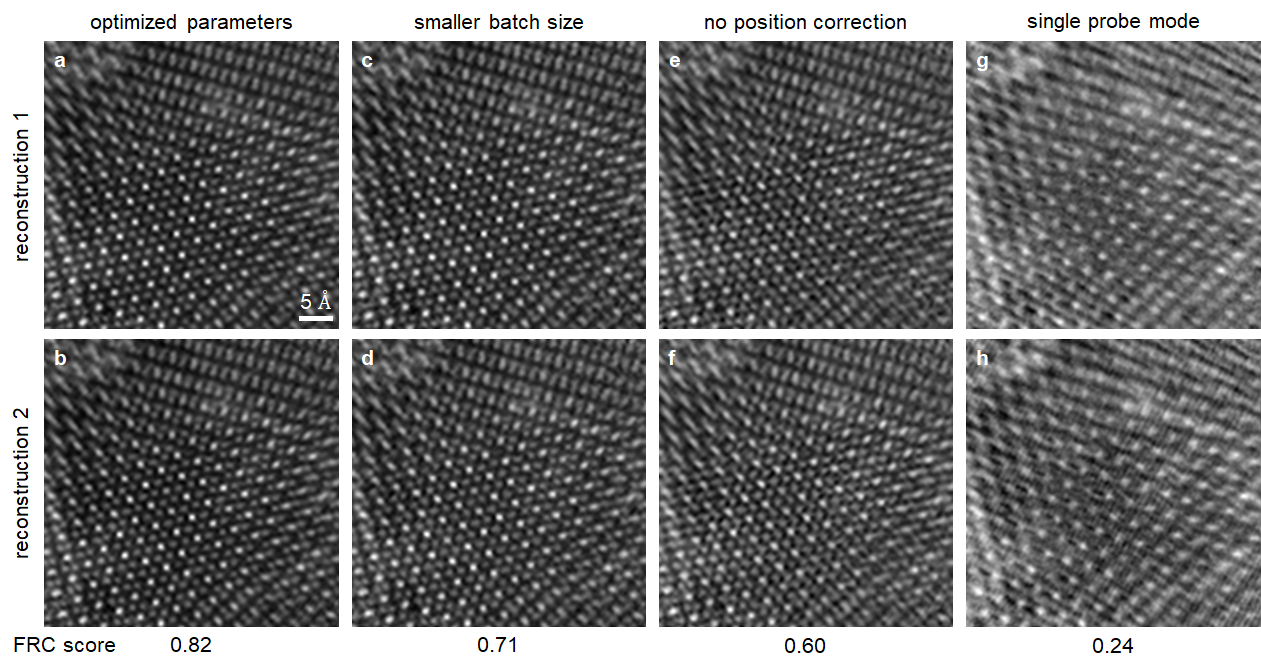}
    \caption{Optimized ptychographic reconstructions of bilayer MoSe$_2$/WS$_2$ compared with sub-optimal parameters. Two datasets that cover the same scan area were reconstructed independently for 5 minutes using the LSQ-ML technique. (a,b) Reconstructed phase with the parameters optimized by Bayesian optimization, including 7 mixed-state probe modes, a batch size of 300, and scan position correction. (c-f) Reconstructions where one of the optimal parameters is changed. All sub-optimal parameter combinations decrease the reconstruction quality to varying degrees. (c,d) Reconstructions with a batch size of 60, which increases the time per iteration, and hence the total number of iterations is reduced. (e,f) Reconstructions without position correction. (g,h) Reconstructions with a single probe mode.}
    \label{fig:recon_param_results}
\end{figure}

Fourier ring correlation (FRC) analysis\cite{harauz1986exact} was used as a quantitative metric to evaluate the quality of ptychographic reconstructions. Without the “ground truth” for experimental data, FRC analysis measures the similarity between two independent reconstructions and is often used to estimate “spatial resolution” in phase retrieval problems\cite{vila2011characterization} or cryogenic electron microscopy reconstructions\cite{van2005fourier}. Our automatic parameter tuning workflow aims to maximize the area under the normalized FRC curve, which ranges from 0 to 1 with 1 corresponding to identical images. The process starts by trying 5 initial random sets of reconstruction parameters, then leverages BO to search for the next point that is most likely to produce better reconstruction quality, and stops after exploring 50 points in total -- only ~1\% of the entire parameter space. A fully random parameter selection strategy was also investigated as comparison. Each sampling strategy was carried out 10 times (with different initial starts) and the averaged best FRC scores after each search are shown in Figure \ref{fig:bo vs random}a. It is obvious that BO can consistently achieve higher FRC scores than random sampling, even if it starts with worse averaged FRC scores. The frequency of specific reconstruction parameter values also demonstrates that BO tends to sample more points around the optimal parameters. For example, for all 3-minute reconstructions, the percentage of position correction used in random and BO sampling are 71.2\% and 48.4\%, respectively. Selected reconstructions from BO and random parameter selections are shown in Figure \ref{fig:bo vs random}b and \ref{fig:bo vs random}d, respectively. The reconstructed atoms from BO parameter tuning are visibly sharper than the ones from random sampling, which agrees with the evaluation based on FRC and their diffractograms (Fourier intensity) (Figure \ref{fig:bo vs random}c\&e). It is worth noting that the diffraction spot that corresponds to an Abbe resolution of 0.42 Å is visible in the reconstruction with optimized parameters, surpassing the resolution (0.69 Å) reported in our previous work\cite{chen2020mixed}.

Figure \ref{fig:recon_param_results} illustrates the importance of reconstruction parameter tuning by showing 5-minute reconstructions. The best results (Figure \ref{fig:recon_param_results}a\&b) found by BO used 7 mixed-states probe modes, a sparse batch size of 300, a Gaussian noise model, and position correction. These parameters agree well with the choice made by experienced scientists who are familiar with the algorithm and the data\cite{chen2020mixed}. For comparison, a smaller batch size of 60 produces reconstructions (Figure \ref{fig:recon_param_results}c\&d) that have broader atoms and a slightly lower FRC score than the optimal results. Moreover, reconstructions with no position correction and only a single probe mode are shown in Figure \ref{fig:recon_param_results}e\&f and g\&h, respectively. Because the experimental errors are not corrected during reconstruction, these images have significantly worse quality and inconsistencies that are reflected in their FRC scores. The complete list of reconstruction parameters used for Figure \ref{fig:recon_param_results} is summarized in Supplementary Table 2.

\begin{figure}
    \centering
    \includegraphics[width = 0.85\linewidth]{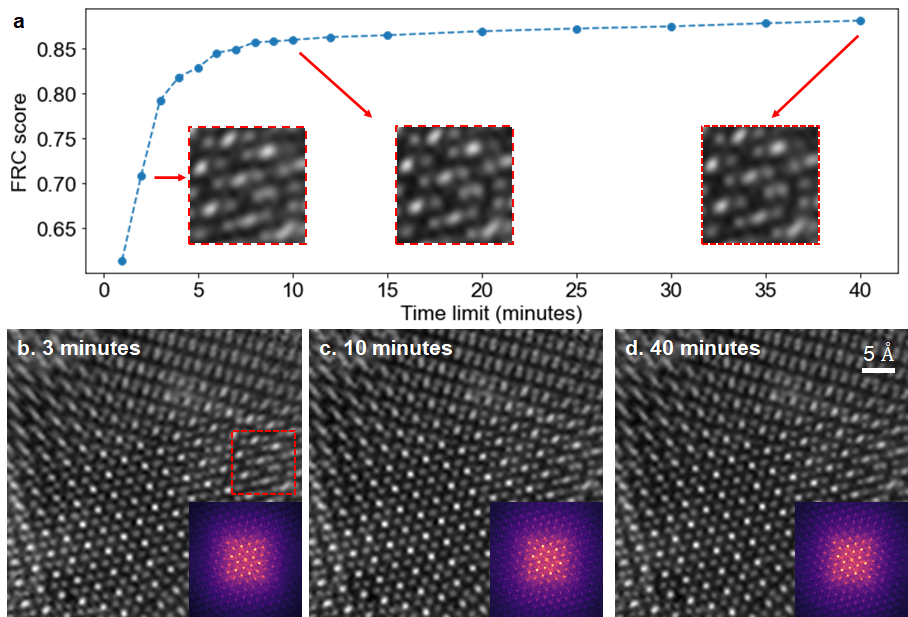}
    \caption{Ptychographic reconstructions of bilayer MoSe$_2$/WS$_2$ at different time limits. The best reconstruction parameters at each time limit were automatically optimized by BO. (a) Best FRC scores vs. reconstruction time. (b-d) Best reconstructions and their diffractograms (insets) after 3, 10, and 40 minutes, respectively. The zoom-ins in (a) shows that reconstructed atoms become more resolved as time increases.}
    \label{fig:fsc_over_time}
\end{figure}

Using the efficient and automatic parameter tuning enabled by BO, one can gain deeper understandings of optimal reconstruction parameters and systematically study how they change with time. As shown in Figure \ref{fig:fsc_over_time}, for the experimental dataset of  MoSe$_2$/WS$_2$, the phase of the object converges to the bilayer structures within 10 minutes, as the probe modes quickly become more physical shapes. As time further increases, the FRC scores continue to improve, which is mainly thanks to  scan position correction that often requires more iterations to refine large drifts or global position errors. The plots for individual reconstruction parameters are provided in Supplementary Figure 1. For all time limits, BO indicates that the best results are obtained with a large number of probe modes, position correction, and Gaussian noise model. It also suggests that no probe variation correction is needed -- this is within our expectations since the scan FOV is relatively small (5 nm $\times$ 5 nm). In the early reconstruction stage ($<$25 minutes), the sparse batch selection scheme 
leads to higher FRC scores since the algorithm has a faster initial convergence rate. On the other hand, given enough time (number of iterations), the algorithm benefits more from the compact batch selection scheme that is known to have slower convergence but is more robust to noise\cite{odstrvcil2018iterative}. We further compared the reconstructed atomic distances with the structure model, and confirmed that the compact batch indeed produces more accurate results than the sparse batch at longer time limits.

\subsection{Optimization of Experimental Parameters for Low-dose Ptychography}

The dose of the illumination beam plays a crucial role in electron microscopy. For example, high electron dose can  damage  the sample structure by energetic electrons, especially for radiation-sensitive samples, such as batteries, metal-organic frameworks, or biological materials\cite{egerton2013control, russo2019damage}. In contrast, low dose mode results in noisy diffraction patterns, reducing spatial resolution or even introducing additional artifacts in ptychographic reconstructions. Therefore, it is critical to explore optimal imaging conditions at the allowed dose of illumination.

In general, experimental conditions, such as scan step size and probe size, often determine the best quality one can achieve after ptychographic reconstruction. The optimal experimental parameters should balance various physical factors such as the signal to noise ratio (SNR) of diffraction patterns, scanning probe overlap, and the sampling requirement in the detector plane\cite{edo2013sampling,zhang2021many}. Due to the complex tradeoffs between different factors, it is generally challenging, even for human experts, to determine the optimal parameters that maximize the reconstruction quality in different experiments. For instance, at a fixed electron dose, a small scan step size leads to a large number of diffraction patterns with electron counts collected in the detector. Increasing scan step size could improve the SNR but reduce the spatial overlap between adjacent probes. Although larger probe defocuses could provide better overlap in real space, it requires higher sampling (more pixels) in the detector plane, which again lowers the SNR since the averaged electron count per pixel is decreased. 

\begin{figure}
    \centering
    \includegraphics[width = 1.0\linewidth]{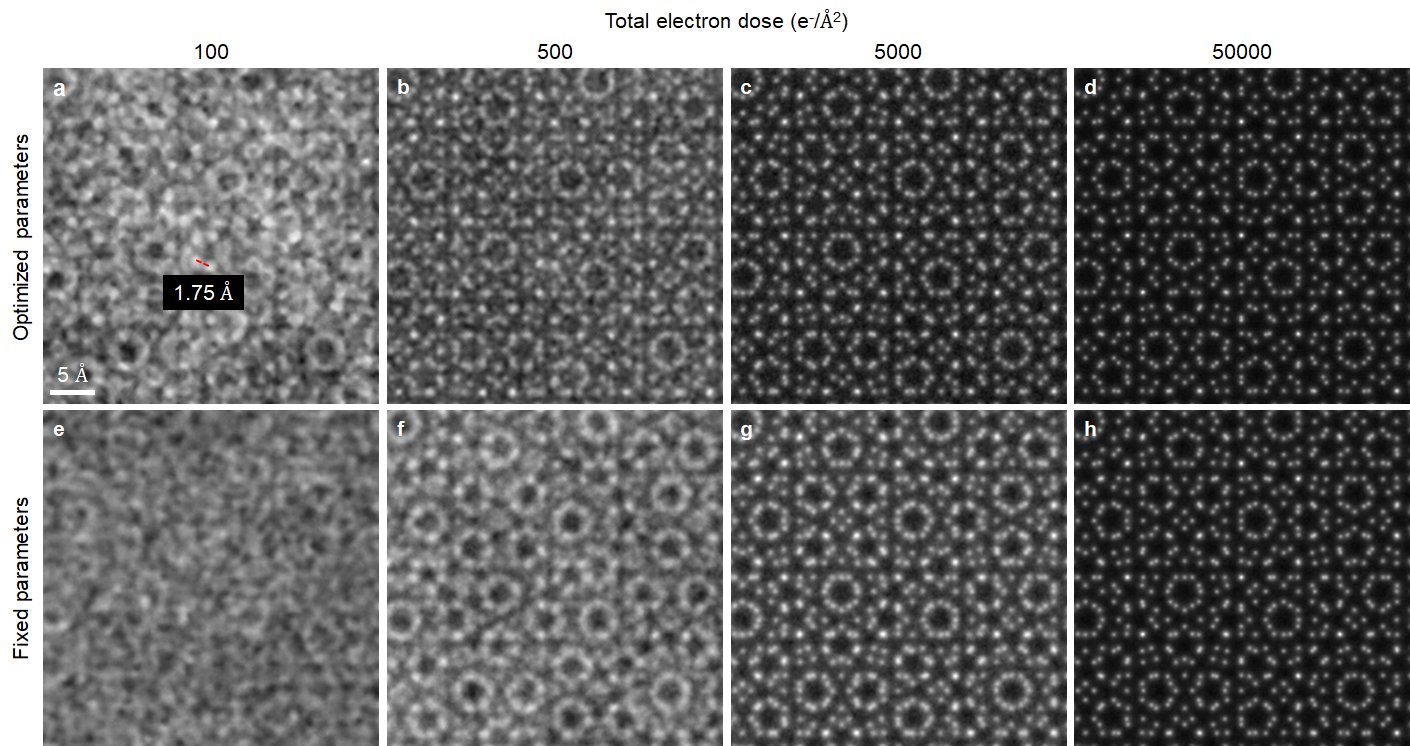}
    \caption{Ptychographic reconstructions of simulated bilayer MoS$_2$ at different total electron doses with an optimized set of experimental parameters compared with an expert-chosen set. (a-d) Phase maps of the reconstructed objects using experimental parameters that are optimized by Bayesian optimization. (e-h) Reconstructions using a fixed set of experimental parameters that are similar to ref.\cite{chen2020mixed}. }
    \label{fig:exp_param_dose}
\end{figure}

Using Bayesian optimization with Gaussian processes, we performed a comprehensive and automatic parameter tuning to search for the optimal scan step size, aperture size, probe defocus, and detector size (Supplementary Table 3) at different electron dose levels. For each point in the 4D parameter space, we first simulated diffraction patterns using a twisted bilayer MoS$_2$ structure (Supplementary Figure 2) as the test object and carried out ptychographic reconstructions using the LSQ-ML algorithm. With the ground truth available, BO directly maximizes the accuracy of reconstruction, which is quantified with the structural similarity index measure (SSIM)\cite{wang2004image}. Figure \ref{fig:exp_param_dose}a-d shows the best reconstructions after 800 points are explored at various dose levels from 100 to 50,000 e$^-$/Å$^2$. As references, reconstructions with a fixed set of experimental parameters (2 Å scan step size, 20 mrad aperture size, -55 nm defocus, 256 $\times$ 256 detector size), which are similar to the ones used in ref. \cite{chen2020mixed}, are shown in Figure \ref{fig:exp_param_dose}e-h. The experimental parameters optimized by BO produce significantly better resolution and more accurate structures, especially at lower dose levels where the physical requirements for good reconstructions are more stringent. At high dose levels, the data have sufficient SNR and the reconstruction quality becomes less sensitive to experimental parameters.

\begin{figure}
    \centering
    \includegraphics[width = 1.0\linewidth]{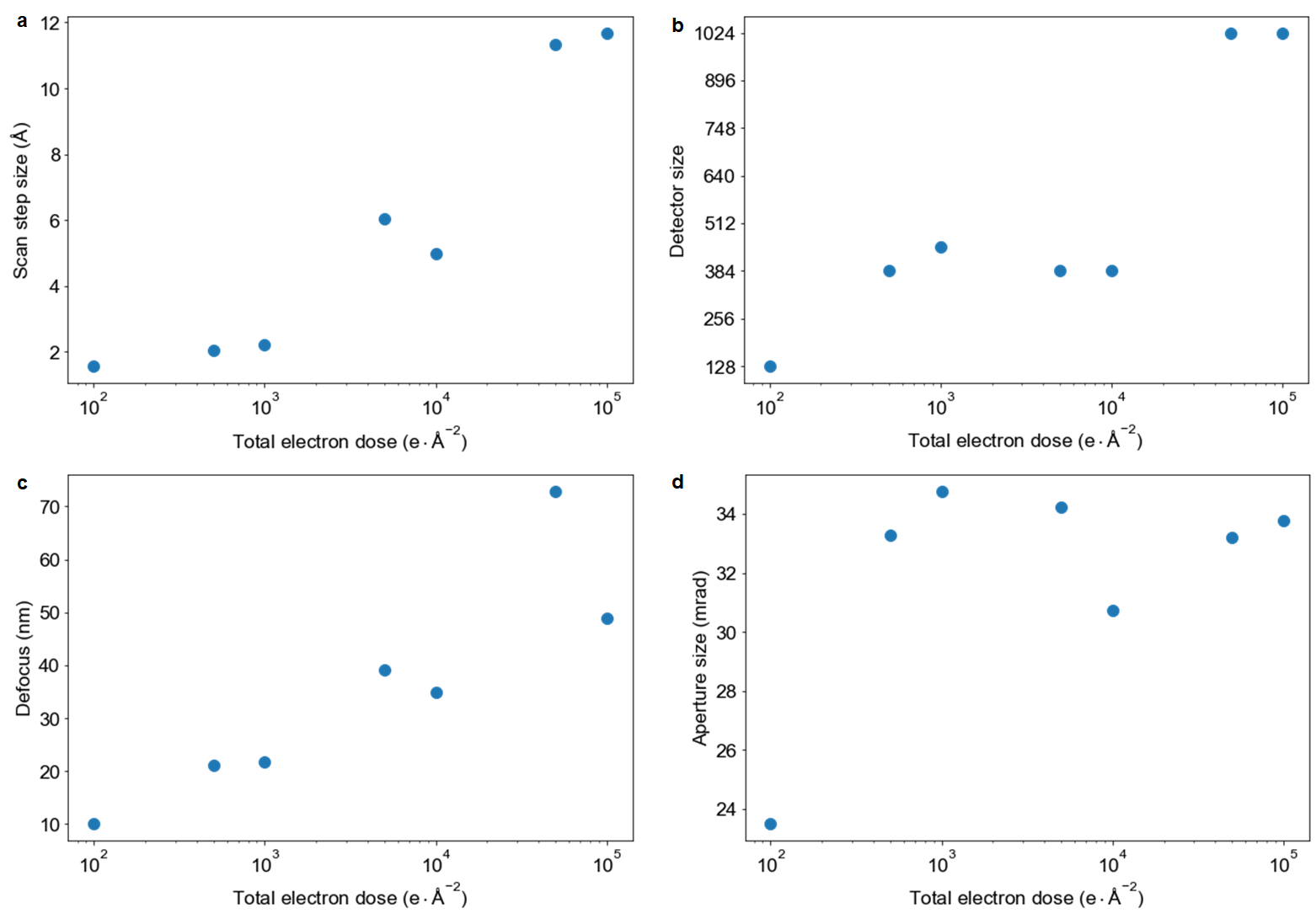}
    \caption{Optimal experimental parameters for electron ptychography. At each electron dose level, BO attempts to optimize scan step size(a), detector size(b), probe defocus(c), and aperture size(d). }
    \label{fig:opt_exp_param}
\end{figure}

The results from BO allowed us to estimate the entire 4D parameter space and observe how optimal experimental conditions depend on the total electron dose. As shown in Figure \ref{fig:opt_exp_param}, small probe and scan step size produce better results at extremely low-dose regimes. However, with increasing total electron counts, one can theoretically use a larger scan step size ($>$5 Å) given sufficient probe overlap and detector pixels. Similarly, it is more advantageous to use a relatively small detector size (e.g. 128 $\times$ 128) at a low dose as more pixels lead to poor SNR. Lastly, with the exception of 100 e$^-$/Å$^2$, most of the optimal conditions found by BO have large ($>$30 mrad) aperture size, indicating that in addition to its size, the probe structure also influences the quality of ptychographic reconstructions. This agrees with previous literature\cite{pelz2017low} that shows specialized focusing optics can produce superior images. Because the focusing probe is typically characterized by a few physical parameters in electron microscopy, we believe the probe structure can be further optimized using the BO framework. This thus provides general guidance for future ptychography experiments.

\section{Discussion}
As a general technique for black-box optimization, BO provides a framework that easily extends to other parameter tuning tasks beyond the eight reconstruction parameters and four experimental parameters studied in this work. For instance, in multi-slice ptychography, one can optimize model parameters such as the sample thickness and the number of layers. By minimizing the data error between reconstruction and data, BO facilitates automatic estimation of experimental conditions, including probe defocus or global scan position errors, which cannot be measured accurately by electron microscopes. In addition, our parameter tuning workflow has implications for similar inverse problems, including tomography or X-ray ptychography. 

Currently, automatic parameter tuning and typical electron ptychography experiments operate on similar time scales (a few hours), which prohibits online parameter tuning. However, multiple hardware and software improvements can be made to further enhance the computational efficiency. First of all, because  the majority of time in the workflow is used for ptychographic reconstruction, utilizing more advanced  hardware, such as high memory-bandwidth GPUs can significantly reduce the processing time by more than tenfold compared to the current work. The total processing time can be further shortened by carrying out multiple reconstructions in parallel and using multi-points optimization strategies\cite{chevalier2013fast}. In addition, experienced scientists may leverage additional properties about reconstruction algorithms or data to reduce the parameter space in BO. The reduction can be implemented at the beginning or during the parameter tuning workflow. Lastly, recent developments such as physics-informed BO\cite{ziatdinov2021physics}, causal BO\cite{aglietti2020causal}, and deep kernel learning\cite{roccapriore2021physics}, may provide more solutions that facilitate more intelligent decision-making by exploiting underlying relationships between different parameters. 

Lastly, we want to emphasize  the importance of the objective function for optimizing reconstruction quality. An ideal objective function should reflect the accuracy of ptychographic reconstructions so that  automatic parameter tuning  produces true sample structures rather than artifacts. In numerical simulations, accuracy can be directly quantified since a ground truth is available. However, for experimental data, the complete information is unknown to researchers and many prevailing evaluation metrics (e.g. the FRC score) only characterize the precision of reconstructions. The FRC analysis is often used in ptychography literature\cite{vila2011characterization,holler2014x,odstrvcil2018iterative,chen2020mixed,jiang2021achieving} and correlates with accuracy to some extent, especially when the dominating factor is dose. Nevertheless, there exist reconstruction and experimental parameters that lead to deceiving results with high precision but low accuracy. For instance, applying image regularization techniques such as de-noising may “improve” the FRC score by removing noisy artifacts in the object, but reduce sharp features if the image is over-smoothed. To avoid such systematic bias, one should be attentive to limitations of different metrics and, if possible, incorporate additional knowledge into the parameter tuning workflow to directly optimize accuracy, or combine with precision measurement via multi-objective optimization\cite{jeong2005efficient}. Future works will explore other precision quantification methods that have better correlation with accuracy as well as simple computational complexity.

\section{Conclusion}
In summary, we demonstrated a human-out-of-loop parameter tuning framework for electron ptychography based on Bayesian optimization with Gaussian processes. The workflow does not require strong prior knowledge about the input data or advanced reconstruction techniques, and can automatically determine the parameters that correctly account for various experimental errors and produce high-resolution ptychographic reconstruction of experimental data. Similarly, BO can be used to search for the optimal experimental conditions in complex multi-dimensional parameter space, allowing better designs for ptychography applications such as low-dose imaging. With rapid developments in computing and control hardware, software, and advanced BO techniques, we anticipate that fully automatic parameter tuning will achieve sufficient throughput for real-time electron ptychography applications.

\section{Acknowledgments}
We thank Wendy Di for helpful discussions. Y.H. and M.C. acknowledge support from the Welch Foundation (C-2065-20210327). This research used resources of the Advanced Photon Source, a U.S. Department of Energy (DOE) Office of Science User Facility operated for the DOE Office of Science by Argonne National Laboratory under Contract No. DE-AC02-06CH11357. 

\section{Methods}

\subsection{Ptychographic reconstruction}
Ptychographic reconstructions were carried out using a customized library based upon the PtychoShelves package\cite{wakonig2020ptychoshelves}. The library, which is maintained at \url{https://github.com/yijiang1/fold_slice}, supports electron ptychography data and provides a python interface. For reconstruction parameter tuning studies, we further modified the code to allow for time-limited reconstruction instead of standard iteration-limited reconstruction.

Supplementary table 1 summarizes eight types of reconstruction parameters that are explored during automatic parameter tuning. These parameters influence both the quality and efficiency of ptychographic reconstruction and are frequently adjusted for different experimental data. The core algorithm is the maximum likelihood ptychography with a least-squares solver\cite{odstrvcil2018iterative}, which provides both Gaussian and Poisson probability distribution to model data noise. The method also used a mini-batch update strategy to efficiently balance reconstruction speed and convergence rate. Thus, the number of diffraction patterns in each batch and the batch selection scheme (sparse vs. compact) are tunable parameters in reconstruction. In addition, the number of probe modes in mixed-states ptychography\cite{thibault2013reconstructing} can be adjusted to account for partial coherence\cite{chen2020mixed} and probe vibration\cite{clark2014dynamic}. In the orthogonal probe relaxation (OPR) technique\cite{odstrcil2016ptychographic}, which is often used to reduce artifacts caused by probe variation in a single scan, the number of orthogonal modes kept in truncated singular value decomposition controls the amount of structural changes at each scan position. Moreover, position correction can refine inaccurate scan positions and intensity correction accounts for changes in probe intensity. Lastly, the "multimodal" option specifies if all or only the first probe mode are used to update the object function.

In general, the upper bounds for the number of mixed-states probe modes, the OPR modes, and the batch size are limited by the data size and the GPU (NVIDIA GeForce GTX 1080 Ti) memory. For simplicity, we define position correction, intensity correction, and multimodal as binary variables. If an option is set to true, then feature is used throughout the entire reconstruction process. 

\subsection{Bayesian Optimization with Gaussian Process}
Bayesian optimization was carried out with the Scikit Optimize library\cite{head2018scikit}. After each ptychographic reconstruction, the image quality and corresponding parameters are used to update the GP model. Here we used the Matern kernel\cite{genton2001classes} -- a popular covariance function defined as:
\begin{align}
    k(x_{i},x_{j}) = \frac{1}{\Gamma(\nu)2^{\nu-1}}\left(\frac{\sqrt{2\nu}}{l}d(x_{i},x_{j})\right)^{\nu}K_{\nu}\left(\frac{\sqrt{2\nu}}{l}d(x_{i},x_{j})\right)
\end{align} 
where $d(x_{i},x_{j})$ measures the Euclidean distance between two points, $\Gamma(\nu)$ is the gamma function, and $K_{\nu}$ is the modified Bessel function of the second kind. $\nu$ is a positive parameter that controls the smoothness of the kernel and $l$ is the length scale, which is updated during BO. For all reconstructions parameter tuning studies in this paper, $\nu$ is set to 1.

To sample the next point, we used a portfolio strategy known as “GP Hedge”\cite{hoffman2011portfolio}, which selects points using a pool of acquisition functions, including negative probability of improvement\cite{kushner1964new}, expected improvement\cite{mockus1978application}, and upper confidence bound\cite{cox1992statistical}.
Lastly, the automatic parameter tuning workflow randomly sample a small number of initial points for Gaussian Process modeling before the Bayesian optimization process.

\subsection{Experimental Parameter Tuning}
For experimental parameter optimization, we generated a simulated potential of bilayer MoS$_2$ with a 30$\degree$ twist. Single-atom potentials\cite{kirkland1998advanced} placed at appropriate coordinates were summed to generate the full potential of the bilayer. Interpolation was used to avoid the large singularity at the center of individual potentials. The resulting potential is 2048 $\times$ 2048 pixels with a pixel size of 0.125 Å.

For all simulated data, the scan field of view was about 6 nm $\times$ 6 nm and the pixel size was fixed at 0.125 Å. The beam energy was set 80 keV. Each 4D dataset was simulated assuming the strong phase approximation, and then reconstructed with a single probe mode, compact batch selection scheme, and no additional corrections. The batch size was chosen dynamically to fully utilize the GPU memory. All reconstructions were run for 500 iterations on a single NVIDIA V100 GPU, and took from $\sim$10 seconds to $\sim$10 minutes, depending on the data size.

As summarized in supplementary Table 2, most experimental parameters are defined as continuous variables, giving an infinite number of possible parameter space. Bayesian optimization attempted to maximize the SSIM\cite{wang2004image} between a reconstruction and the ground truth. The Hammersley sampling method\cite{hammersley1960monte} was used to explore 100 initial points that randomly cover the entire parameter, after which BO was used to search for additional 700 sets of parameters.

\section{Availability of Data and Materials}
Diffraction data used for reconstruction parameter tuning are included in this published article by Chen et al.\cite{chen2020mixed}. All other data and code are available from the corresponding author at reasonable request.
\bibliography{mybib.bib}


\begin{thebibliography}{52}
\ifx \bisbn   \undefined \def \bisbn  #1{ISBN #1}\fi
\ifx \binits  \undefined \def \binits#1{#1}\fi
\ifx \bauthor  \undefined \def \bauthor#1{#1}\fi
\ifx \batitle  \undefined \def \batitle#1{#1}\fi
\ifx \bjtitle  \undefined \def \bjtitle#1{#1}\fi
\ifx \bvolume  \undefined \def \bvolume#1{\textbf{#1}}\fi
\ifx \byear  \undefined \def \byear#1{#1}\fi
\ifx \bissue  \undefined \def \bissue#1{#1}\fi
\ifx \bfpage  \undefined \def \bfpage#1{#1}\fi
\ifx \blpage  \undefined \def \blpage #1{#1}\fi
\ifx \burl  \undefined \def \burl#1{\textsf{#1}}\fi
\ifx \doiurl  \undefined \def \doiurl#1{\url{https://doi.org/#1}}\fi
\ifx \betal  \undefined \def \betal{\textit{et al.}}\fi
\ifx \binstitute  \undefined \def \binstitute#1{#1}\fi
\ifx \binstitutionaled  \undefined \def \binstitutionaled#1{#1}\fi
\ifx \bctitle  \undefined \def \bctitle#1{#1}\fi
\ifx \beditor  \undefined \def \beditor#1{#1}\fi
\ifx \bpublisher  \undefined \def \bpublisher#1{#1}\fi
\ifx \bbtitle  \undefined \def \bbtitle#1{#1}\fi
\ifx \bedition  \undefined \def \bedition#1{#1}\fi
\ifx \bseriesno  \undefined \def \bseriesno#1{#1}\fi
\ifx \blocation  \undefined \def \blocation#1{#1}\fi
\ifx \bsertitle  \undefined \def \bsertitle#1{#1}\fi
\ifx \bsnm \undefined \def \bsnm#1{#1}\fi
\ifx \bsuffix \undefined \def \bsuffix#1{#1}\fi
\ifx \bparticle \undefined \def \bparticle#1{#1}\fi
\ifx \barticle \undefined \def \barticle#1{#1}\fi
\bibcommenthead
\ifx \bconfdate \undefined \def \bconfdate #1{#1}\fi
\ifx \botherref \undefined \def \botherref #1{#1}\fi
\ifx \url \undefined \def \url#1{\textsf{#1}}\fi
\ifx \bchapter \undefined \def \bchapter#1{#1}\fi
\ifx \bbook \undefined \def \bbook#1{#1}\fi
\ifx \bcomment \undefined \def \bcomment#1{#1}\fi
\ifx \oauthor \undefined \def \oauthor#1{#1}\fi
\ifx \citeauthoryear \undefined \def \citeauthoryear#1{#1}\fi
\ifx \endbibitem  \undefined \def \endbibitem {}\fi
\ifx \bconflocation  \undefined \def \bconflocation#1{#1}\fi
\ifx \arxivurl  \undefined \def \arxivurl#1{\textsf{#1}}\fi
\csname PreBibitemsHook\endcsname

\bibitem{yang2016simultaneous}
\begin{barticle}
\bauthor{\bsnm{Yang}, \binits{H.}},
\bauthor{\bsnm{Rutte}, \binits{R.}},
\bauthor{\bsnm{Jones}, \binits{L.}},
\bauthor{\bsnm{Simson}, \binits{M.}},
\bauthor{\bsnm{Sagawa}, \binits{R.}},
\bauthor{\bsnm{Ryll}, \binits{H.}},
\bauthor{\bsnm{Huth}, \binits{M.}},
\bauthor{\bsnm{Pennycook}, \binits{T.}},
\bauthor{\bsnm{Green}, \binits{M.}},
\bauthor{\bsnm{Soltau}, \binits{H.}}, \betal:
\batitle{Simultaneous atomic-resolution electron ptychography and z-contrast
  imaging of light and heavy elements in complex nanostructures}.
\bjtitle{Nature Communications}
\bvolume{7}(\bissue{1}),
\bfpage{1}--\blpage{8}
(\byear{2016})
\end{barticle}
\endbibitem

\bibitem{gao2017electron}
\begin{barticle}
\bauthor{\bsnm{Gao}, \binits{S.}},
\bauthor{\bsnm{Wang}, \binits{P.}},
\bauthor{\bsnm{Zhang}, \binits{F.}},
\bauthor{\bsnm{Martinez}, \binits{G.T.}},
\bauthor{\bsnm{Nellist}, \binits{P.D.}},
\bauthor{\bsnm{Pan}, \binits{X.}},
\bauthor{\bsnm{Kirkland}, \binits{A.I.}}:
\batitle{Electron ptychographic microscopy for three-dimensional imaging}.
\bjtitle{Nature communications}
\bvolume{8}(\bissue{1}),
\bfpage{1}--\blpage{8}
(\byear{2017})
\end{barticle}
\endbibitem

\bibitem{jiang2018electron}
\begin{barticle}
\bauthor{\bsnm{Jiang}, \binits{Y.}},
\bauthor{\bsnm{Chen}, \binits{Z.}},
\bauthor{\bsnm{Han}, \binits{Y.}},
\bauthor{\bsnm{Deb}, \binits{P.}},
\bauthor{\bsnm{Gao}, \binits{H.}},
\bauthor{\bsnm{Xie}, \binits{S.}},
\bauthor{\bsnm{Purohit}, \binits{P.}},
\bauthor{\bsnm{Tate}, \binits{M.W.}},
\bauthor{\bsnm{Park}, \binits{J.}},
\bauthor{\bsnm{Gruner}, \binits{S.M.}}, \betal:
\batitle{Electron ptychography of 2d materials to deep sub-{\aa}ngstr{\"o}m
  resolution}.
\bjtitle{Nature}
\bvolume{559}(\bissue{7714}),
\bfpage{343}--\blpage{349}
(\byear{2018})
\end{barticle}
\endbibitem

\bibitem{chen2021electron}
\begin{barticle}
\bauthor{\bsnm{Chen}, \binits{Z.}},
\bauthor{\bsnm{Jiang}, \binits{Y.}},
\bauthor{\bsnm{Shao}, \binits{Y.-T.}},
\bauthor{\bsnm{Holtz}, \binits{M.E.}},
\bauthor{\bsnm{Odstr{\v{c}}il}, \binits{M.}},
\bauthor{\bsnm{Guizar-Sicairos}, \binits{M.}},
\bauthor{\bsnm{Hanke}, \binits{I.}},
\bauthor{\bsnm{Ganschow}, \binits{S.}},
\bauthor{\bsnm{Schlom}, \binits{D.G.}},
\bauthor{\bsnm{Muller}, \binits{D.A.}}:
\batitle{Electron ptychography achieves atomic-resolution limits set by lattice
  vibrations}.
\bjtitle{Science}
\bvolume{372}(\bissue{6544}),
\bfpage{826}--\blpage{831}
(\byear{2021})
\end{barticle}
\endbibitem

\bibitem{hoppe1969beugung}
\begin{barticle}
\bauthor{\bsnm{Hoppe}, \binits{W.}}:
\batitle{Beugung im inhomogenen prim{\"a}rstrahlwellenfeld. i. prinzip einer
  phasenmessung von elektronenbeungungsinterferenzen}.
\bjtitle{Acta Crystallographica Section A: Crystal Physics, Diffraction,
  Theoretical and General Crystallography}
\bvolume{25}(\bissue{4}),
\bfpage{495}--\blpage{501}
(\byear{1969})
\end{barticle}
\endbibitem

\bibitem{ophus2014recording}
\begin{barticle}
\bauthor{\bsnm{Ophus}, \binits{C.}},
\bauthor{\bsnm{Ercius}, \binits{P.}},
\bauthor{\bsnm{Sarahan}, \binits{M.}},
\bauthor{\bsnm{Czarnik}, \binits{C.}},
\bauthor{\bsnm{Ciston}, \binits{J.}}:
\batitle{Recording and using 4d-stem datasets in materials science}.
\bjtitle{Microscopy and Microanalysis}
\bvolume{20}(\bissue{S3}),
\bfpage{62}--\blpage{63}
(\byear{2014})
\end{barticle}
\endbibitem

\bibitem{tate2016high}
\begin{barticle}
\bauthor{\bsnm{Tate}, \binits{M.W.}},
\bauthor{\bsnm{Purohit}, \binits{P.}},
\bauthor{\bsnm{Chamberlain}, \binits{D.}},
\bauthor{\bsnm{Nguyen}, \binits{K.X.}},
\bauthor{\bsnm{Hovden}, \binits{R.}},
\bauthor{\bsnm{Chang}, \binits{C.S.}},
\bauthor{\bsnm{Deb}, \binits{P.}},
\bauthor{\bsnm{Turgut}, \binits{E.}},
\bauthor{\bsnm{Heron}, \binits{J.T.}},
\bauthor{\bsnm{Schlom}, \binits{D.G.}}, \betal:
\batitle{High dynamic range pixel array detector for scanning transmission
  electron microscopy}.
\bjtitle{Microscopy and Microanalysis}
\bvolume{22}(\bissue{1}),
\bfpage{237}--\blpage{249}
(\byear{2016})
\end{barticle}
\endbibitem

\bibitem{mir2017characterisation}
\begin{barticle}
\bauthor{\bsnm{Mir}, \binits{J.}},
\bauthor{\bsnm{Clough}, \binits{R.}},
\bauthor{\bsnm{MacInnes}, \binits{R.}},
\bauthor{\bsnm{Gough}, \binits{C.}},
\bauthor{\bsnm{Plackett}, \binits{R.}},
\bauthor{\bsnm{Shipsey}, \binits{I.}},
\bauthor{\bsnm{Sawada}, \binits{H.}},
\bauthor{\bsnm{MacLaren}, \binits{I.}},
\bauthor{\bsnm{Ballabriga}, \binits{R.}},
\bauthor{\bsnm{Maneuski}, \binits{D.}}, \betal:
\batitle{Characterisation of the medipix3 detector for 60 and 80 kev
  electrons}.
\bjtitle{Ultramicroscopy}
\bvolume{182},
\bfpage{44}--\blpage{53}
(\byear{2017})
\end{barticle}
\endbibitem

\bibitem{thibault2008high}
\begin{barticle}
\bauthor{\bsnm{Thibault}, \binits{P.}},
\bauthor{\bsnm{Dierolf}, \binits{M.}},
\bauthor{\bsnm{Menzel}, \binits{A.}},
\bauthor{\bsnm{Bunk}, \binits{O.}},
\bauthor{\bsnm{David}, \binits{C.}},
\bauthor{\bsnm{Pfeiffer}, \binits{F.}}:
\batitle{High-resolution scanning x-ray diffraction microscopy}.
\bjtitle{Science}
\bvolume{321}(\bissue{5887}),
\bfpage{379}--\blpage{382}
(\byear{2008})
\end{barticle}
\endbibitem

\bibitem{guizar2008phase}
\begin{barticle}
\bauthor{\bsnm{Guizar-Sicairos}, \binits{M.}},
\bauthor{\bsnm{Fienup}, \binits{J.R.}}:
\batitle{Phase retrieval with transverse translation diversity: a nonlinear
  optimization approach}.
\bjtitle{Optics express}
\bvolume{16}(\bissue{10}),
\bfpage{7264}--\blpage{7278}
(\byear{2008})
\end{barticle}
\endbibitem

\bibitem{maiden2009improved}
\begin{barticle}
\bauthor{\bsnm{Maiden}, \binits{A.M.}},
\bauthor{\bsnm{Rodenburg}, \binits{J.M.}}:
\batitle{An improved ptychographical phase retrieval algorithm for diffractive
  imaging}.
\bjtitle{Ultramicroscopy}
\bvolume{109}(\bissue{10}),
\bfpage{1256}--\blpage{1262}
(\byear{2009})
\end{barticle}
\endbibitem

\bibitem{maiden2011superresolution}
\begin{barticle}
\bauthor{\bsnm{Maiden}, \binits{A.M.}},
\bauthor{\bsnm{Humphry}, \binits{M.J.}},
\bauthor{\bsnm{Zhang}, \binits{F.}},
\bauthor{\bsnm{Rodenburg}, \binits{J.M.}}:
\batitle{Superresolution imaging via ptychography}.
\bjtitle{JOSA A}
\bvolume{28}(\bissue{4}),
\bfpage{604}--\blpage{612}
(\byear{2011})
\end{barticle}
\endbibitem

\bibitem{chen2020mixed}
\begin{barticle}
\bauthor{\bsnm{Chen}, \binits{Z.}},
\bauthor{\bsnm{Odstrcil}, \binits{M.}},
\bauthor{\bsnm{Jiang}, \binits{Y.}},
\bauthor{\bsnm{Han}, \binits{Y.}},
\bauthor{\bsnm{Chiu}, \binits{M.-H.}},
\bauthor{\bsnm{Li}, \binits{L.-J.}},
\bauthor{\bsnm{Muller}, \binits{D.A.}}:
\batitle{Mixed-state electron ptychography enables sub-angstrom resolution
  imaging with picometer precision at low dose}.
\bjtitle{Nature communications}
\bvolume{11}(\bissue{1}),
\bfpage{1}--\blpage{10}
(\byear{2020})
\end{barticle}
\endbibitem

\bibitem{song2019atomic}
\begin{barticle}
\bauthor{\bsnm{Song}, \binits{J.}},
\bauthor{\bsnm{Allen}, \binits{C.S.}},
\bauthor{\bsnm{Gao}, \binits{S.}},
\bauthor{\bsnm{Huang}, \binits{C.}},
\bauthor{\bsnm{Sawada}, \binits{H.}},
\bauthor{\bsnm{Pan}, \binits{X.}},
\bauthor{\bsnm{Warner}, \binits{J.}},
\bauthor{\bsnm{Wang}, \binits{P.}},
\bauthor{\bsnm{Kirkland}, \binits{A.I.}}:
\batitle{Atomic resolution defocused electron ptychography at low dose with a
  fast, direct electron detector}.
\bjtitle{Scientific reports}
\bvolume{9}(\bissue{1}),
\bfpage{1}--\blpage{8}
(\byear{2019})
\end{barticle}
\endbibitem

\bibitem{pelz2017low}
\begin{barticle}
\bauthor{\bsnm{Pelz}, \binits{P.M.}},
\bauthor{\bsnm{Qiu}, \binits{W.X.}},
\bauthor{\bsnm{B{\"u}cker}, \binits{R.}},
\bauthor{\bsnm{Kassier}, \binits{G.}},
\bauthor{\bsnm{Miller}, \binits{R.D.}}:
\batitle{Low-dose cryo electron ptychography via non-convex bayesian
  optimization}.
\bjtitle{Scientific reports}
\bvolume{7}(\bissue{1}),
\bfpage{1}--\blpage{13}
(\byear{2017})
\end{barticle}
\endbibitem

\bibitem{zhou2020low}
\begin{barticle}
\bauthor{\bsnm{Zhou}, \binits{L.}},
\bauthor{\bsnm{Song}, \binits{J.}},
\bauthor{\bsnm{Kim}, \binits{J.S.}},
\bauthor{\bsnm{Pei}, \binits{X.}},
\bauthor{\bsnm{Huang}, \binits{C.}},
\bauthor{\bsnm{Boyce}, \binits{M.}},
\bauthor{\bsnm{Mendon{\c{c}}a}, \binits{L.}},
\bauthor{\bsnm{Clare}, \binits{D.}},
\bauthor{\bsnm{Siebert}, \binits{A.}},
\bauthor{\bsnm{Allen}, \binits{C.S.}}, \betal:
\batitle{Low-dose phase retrieval of biological specimens using cryo-electron
  ptychography}.
\bjtitle{Nature communications}
\bvolume{11}(\bissue{1}),
\bfpage{1}--\blpage{9}
(\byear{2020})
\end{barticle}
\endbibitem

\bibitem{chen2021three}
\begin{barticle}
\bauthor{\bsnm{Chen}, \binits{Z.}},
\bauthor{\bsnm{Shao}, \binits{Y.-T.}},
\bauthor{\bsnm{Jiang}, \binits{Y.}},
\bauthor{\bsnm{Muller}, \binits{D.}}:
\batitle{Three-dimensional imaging of single dopants inside crystals using
  multislice electron ptychography}.
\bjtitle{Microscopy and Microanalysis}
\bvolume{27}(\bissue{S1}),
\bfpage{2146}--\blpage{2148}
(\byear{2021})
\end{barticle}
\endbibitem

\bibitem{rasmussen2003gaussian}
\begin{bchapter}
\bauthor{\bsnm{Rasmussen}, \binits{C.E.}}:
\bctitle{Gaussian processes in machine learning}.
In: \bbtitle{Summer School on Machine Learning},
pp. \bfpage{63}--\blpage{71}
(\byear{2003}).
\bcomment{Springer}
\end{bchapter}
\endbibitem

\bibitem{bergstra2011algorithms}
\begin{botherref}
\oauthor{\bsnm{Bergstra}, \binits{J.}},
\oauthor{\bsnm{Bardenet}, \binits{R.}},
\oauthor{\bsnm{Bengio}, \binits{Y.}},
\oauthor{\bsnm{K{\'e}gl}, \binits{B.}}:
Algorithms for hyper-parameter optimization.
Advances in neural information processing systems
\textbf{24}
(2011)
\end{botherref}
\endbibitem

\bibitem{brochu2010tutorial}
\begin{botherref}
\oauthor{\bsnm{Brochu}, \binits{E.}},
\oauthor{\bsnm{Cora}, \binits{V.M.}},
\oauthor{\bsnm{De~Freitas}, \binits{N.}}:
A tutorial on bayesian optimization of expensive cost functions, with
  application to active user modeling and hierarchical reinforcement learning.
arXiv preprint arXiv:1012.2599
(2010)
\end{botherref}
\endbibitem

\bibitem{mahendran2012adaptive}
\begin{bchapter}
\bauthor{\bsnm{Mahendran}, \binits{N.}},
\bauthor{\bsnm{Wang}, \binits{Z.}},
\bauthor{\bsnm{Hamze}, \binits{F.}},
\bauthor{\bsnm{De~Freitas}, \binits{N.}}:
\bctitle{Adaptive mcmc with bayesian optimization}.
In: \bbtitle{Artificial Intelligence and Statistics},
pp. \bfpage{751}--\blpage{760}
(\byear{2012}).
\bcomment{PMLR}
\end{bchapter}
\endbibitem

\bibitem{duris2020bayesian}
\begin{barticle}
\bauthor{\bsnm{Duris}, \binits{J.}},
\bauthor{\bsnm{Kennedy}, \binits{D.}},
\bauthor{\bsnm{Hanuka}, \binits{A.}},
\bauthor{\bsnm{Shtalenkova}, \binits{J.}},
\bauthor{\bsnm{Edelen}, \binits{A.}},
\bauthor{\bsnm{Baxevanis}, \binits{P.}},
\bauthor{\bsnm{Egger}, \binits{A.}},
\bauthor{\bsnm{Cope}, \binits{T.}},
\bauthor{\bsnm{McIntire}, \binits{M.}},
\bauthor{\bsnm{Ermon}, \binits{S.}}, \betal:
\batitle{Bayesian optimization of a free-electron laser}.
\bjtitle{Physical review letters}
\bvolume{124}(\bissue{12}),
\bfpage{124801}
(\byear{2020})
\end{barticle}
\endbibitem

\bibitem{zhang2021aberration}
\begin{barticle}
\bauthor{\bsnm{Zhang}, \binits{C.}},
\bauthor{\bsnm{Baraissov}, \binits{Z.}},
\bauthor{\bsnm{Duncan}, \binits{C.}},
\bauthor{\bsnm{Hanuka}, \binits{A.}},
\bauthor{\bsnm{Edelen}, \binits{A.}},
\bauthor{\bsnm{Maxson}, \binits{J.}},
\bauthor{\bsnm{Muller}, \binits{D.}}:
\batitle{Aberration corrector tuning with machine-learning-based emittance
  measurements and bayesian optimization}.
\bjtitle{Microscopy and Microanalysis}
\bvolume{27}(\bissue{S1}),
\bfpage{810}--\blpage{812}
(\byear{2021})
\end{barticle}
\endbibitem

\bibitem{roccapriore2021physics}
\begin{botherref}
\oauthor{\bsnm{Roccapriore}, \binits{K.M.}},
\oauthor{\bsnm{Kalinin}, \binits{S.V.}},
\oauthor{\bsnm{Ziatdinov}, \binits{M.}}:
Physics discovery in nanoplasmonic systems via autonomous experiments in
  scanning transmission electron microscopy.
arXiv preprint arXiv:2108.03290
(2021)
\end{botherref}
\endbibitem

\bibitem{odstrvcil2018iterative}
\begin{barticle}
\bauthor{\bsnm{Odstr{\v{c}}il}, \binits{M.}},
\bauthor{\bsnm{Menzel}, \binits{A.}},
\bauthor{\bsnm{Guizar-Sicairos}, \binits{M.}}:
\batitle{Iterative least-squares solver for generalized maximum-likelihood
  ptychography}.
\bjtitle{Optics express}
\bvolume{26}(\bissue{3}),
\bfpage{3108}--\blpage{3123}
(\byear{2018})
\end{barticle}
\endbibitem

\bibitem{wakonig2020ptychoshelves}
\begin{barticle}
\bauthor{\bsnm{Wakonig}, \binits{K.}},
\bauthor{\bsnm{Stadler}, \binits{H.-C.}},
\bauthor{\bsnm{Odstr{\v{c}}il}, \binits{M.}},
\bauthor{\bsnm{Tsai}, \binits{E.H.}},
\bauthor{\bsnm{Diaz}, \binits{A.}},
\bauthor{\bsnm{Holler}, \binits{M.}},
\bauthor{\bsnm{Usov}, \binits{I.}},
\bauthor{\bsnm{Raabe}, \binits{J.}},
\bauthor{\bsnm{Menzel}, \binits{A.}},
\bauthor{\bsnm{Guizar-Sicairos}, \binits{M.}}:
\batitle{Ptychoshelves, a versatile high-level framework for high-performance
  analysis of ptychographic data}.
\bjtitle{Journal of applied crystallography}
\bvolume{53}(\bissue{2}),
\bfpage{574}--\blpage{586}
(\byear{2020})
\end{barticle}
\endbibitem

\bibitem{thibault2013reconstructing}
\begin{barticle}
\bauthor{\bsnm{Thibault}, \binits{P.}},
\bauthor{\bsnm{Menzel}, \binits{A.}}:
\batitle{Reconstructing state mixtures from diffraction measurements}.
\bjtitle{Nature}
\bvolume{494}(\bissue{7435}),
\bfpage{68}--\blpage{71}
(\byear{2013})
\end{barticle}
\endbibitem

\bibitem{odstrcil2016ptychographic}
\begin{barticle}
\bauthor{\bsnm{Odstrcil}, \binits{M.}},
\bauthor{\bsnm{Baksh}, \binits{P.}},
\bauthor{\bsnm{Boden}, \binits{S.}},
\bauthor{\bsnm{Card}, \binits{R.}},
\bauthor{\bsnm{Chad}, \binits{J.}},
\bauthor{\bsnm{Frey}, \binits{J.}},
\bauthor{\bsnm{Brocklesby}, \binits{W.}}:
\batitle{Ptychographic coherent diffractive imaging with orthogonal probe
  relaxation}.
\bjtitle{Optics express}
\bvolume{24}(\bissue{8}),
\bfpage{8360}--\blpage{8369}
(\byear{2016})
\end{barticle}
\endbibitem

\bibitem{tsai2016x}
\begin{barticle}
\bauthor{\bsnm{Tsai}, \binits{E.H.}},
\bauthor{\bsnm{Usov}, \binits{I.}},
\bauthor{\bsnm{Diaz}, \binits{A.}},
\bauthor{\bsnm{Menzel}, \binits{A.}},
\bauthor{\bsnm{Guizar-Sicairos}, \binits{M.}}:
\batitle{X-ray ptychography with extended depth of field}.
\bjtitle{Optics express}
\bvolume{24}(\bissue{25}),
\bfpage{29089}--\blpage{29108}
(\byear{2016})
\end{barticle}
\endbibitem

\bibitem{jiang2021achieving}
\begin{barticle}
\bauthor{\bsnm{Jiang}, \binits{Y.}},
\bauthor{\bsnm{Deng}, \binits{J.}},
\bauthor{\bsnm{Yao}, \binits{Y.}},
\bauthor{\bsnm{Klug}, \binits{J.A.}},
\bauthor{\bsnm{Mashrafi}, \binits{S.}},
\bauthor{\bsnm{Roehrig}, \binits{C.}},
\bauthor{\bsnm{Preissner}, \binits{C.}},
\bauthor{\bsnm{Marin}, \binits{F.S.}},
\bauthor{\bsnm{Cai}, \binits{Z.}},
\bauthor{\bsnm{Lai}, \binits{B.}}, \betal:
\batitle{Achieving high spatial resolution in a large field-of-view using
  lensless x-ray imaging}.
\bjtitle{Applied Physics Letters}
\bvolume{119}(\bissue{12}),
\bfpage{124101}
(\byear{2021})
\end{barticle}
\endbibitem

\bibitem{harauz1986exact}
\begin{barticle}
\bauthor{\bsnm{Harauz}, \binits{G.}},
\bauthor{\bparticle{van} \bsnm{Heel}, \binits{M.}}:
\batitle{Exact filters for general geometry three dimensional reconstruction.}
\bjtitle{Optik.}
\bvolume{73}(\bissue{4}),
\bfpage{146}--\blpage{156}
(\byear{1986})
\end{barticle}
\endbibitem

\bibitem{vila2011characterization}
\begin{barticle}
\bauthor{\bsnm{Vila-Comamala}, \binits{J.}},
\bauthor{\bsnm{Diaz}, \binits{A.}},
\bauthor{\bsnm{Guizar-Sicairos}, \binits{M.}},
\bauthor{\bsnm{Mantion}, \binits{A.}},
\bauthor{\bsnm{Kewish}, \binits{C.M.}},
\bauthor{\bsnm{Menzel}, \binits{A.}},
\bauthor{\bsnm{Bunk}, \binits{O.}},
\bauthor{\bsnm{David}, \binits{C.}}:
\batitle{Characterization of high-resolution diffractive x-ray optics by
  ptychographic coherent diffractive imaging}.
\bjtitle{Optics express}
\bvolume{19}(\bissue{22}),
\bfpage{21333}--\blpage{21344}
(\byear{2011})
\end{barticle}
\endbibitem

\bibitem{van2005fourier}
\begin{barticle}
\bauthor{\bsnm{Van~Heel}, \binits{M.}},
\bauthor{\bsnm{Schatz}, \binits{M.}}:
\batitle{Fourier shell correlation threshold criteria}.
\bjtitle{Journal of structural biology}
\bvolume{151}(\bissue{3}),
\bfpage{250}--\blpage{262}
(\byear{2005})
\end{barticle}
\endbibitem

\bibitem{egerton2013control}
\begin{barticle}
\bauthor{\bsnm{Egerton}, \binits{R.}}:
\batitle{Control of radiation damage in the tem}.
\bjtitle{Ultramicroscopy}
\bvolume{127},
\bfpage{100}--\blpage{108}
(\byear{2013})
\end{barticle}
\endbibitem

\bibitem{russo2019damage}
\begin{barticle}
\bauthor{\bsnm{Russo}, \binits{C.}},
\bauthor{\bsnm{Egerton}, \binits{R.}}:
\batitle{Damage in electron cryomicroscopy: Lessons from biology for materials
  science}.
\bjtitle{MRS Bulletin}
\bvolume{44}(\bissue{12}),
\bfpage{935}--\blpage{941}
(\byear{2019})
\end{barticle}
\endbibitem

\bibitem{edo2013sampling}
\begin{barticle}
\bauthor{\bsnm{Edo}, \binits{T.}},
\bauthor{\bsnm{Batey}, \binits{D.}},
\bauthor{\bsnm{Maiden}, \binits{A.}},
\bauthor{\bsnm{Rau}, \binits{C.}},
\bauthor{\bsnm{Wagner}, \binits{U.}},
\bauthor{\bsnm{Pe{\v{s}}i{\'c}}, \binits{Z.}},
\bauthor{\bsnm{Waigh}, \binits{T.}},
\bauthor{\bsnm{Rodenburg}, \binits{J.}}:
\batitle{Sampling in x-ray ptychography}.
\bjtitle{Physical Review A}
\bvolume{87}(\bissue{5}),
\bfpage{053850}
(\byear{2013})
\end{barticle}
\endbibitem

\bibitem{zhang2021many}
\begin{barticle}
\bauthor{\bsnm{Zhang}, \binits{X.}},
\bauthor{\bsnm{Chen}, \binits{Z.}},
\bauthor{\bsnm{Muller}, \binits{D.}}:
\batitle{How many detector pixels do we need for super-resolution
  ptychography?}
\bjtitle{Microscopy and Microanalysis}
\bvolume{27}(\bissue{S1}),
\bfpage{620}--\blpage{622}
(\byear{2021})
\end{barticle}
\endbibitem

\bibitem{wang2004image}
\begin{barticle}
\bauthor{\bsnm{Wang}, \binits{Z.}},
\bauthor{\bsnm{Bovik}, \binits{A.C.}},
\bauthor{\bsnm{Sheikh}, \binits{H.R.}},
\bauthor{\bsnm{Simoncelli}, \binits{E.P.}}:
\batitle{Image quality assessment: from error visibility to structural
  similarity}.
\bjtitle{IEEE transactions on image processing}
\bvolume{13}(\bissue{4}),
\bfpage{600}--\blpage{612}
(\byear{2004})
\end{barticle}
\endbibitem

\bibitem{chevalier2013fast}
\begin{bchapter}
\bauthor{\bsnm{Chevalier}, \binits{C.}},
\bauthor{\bsnm{Ginsbourger}, \binits{D.}}:
\bctitle{Fast computation of the multi-points expected improvement with
  applications in batch selection}.
In: \bbtitle{International Conference on Learning and Intelligent
  Optimization},
pp. \bfpage{59}--\blpage{69}
(\byear{2013}).
\bcomment{Springer}
\end{bchapter}
\endbibitem

\bibitem{ziatdinov2021physics}
\begin{botherref}
\oauthor{\bsnm{Ziatdinov}, \binits{M.}},
\oauthor{\bsnm{Ghosh}, \binits{A.}},
\oauthor{\bsnm{Kalinin}, \binits{S.V.}}:
Physics makes the difference: Bayesian optimization and active learning via
  augmented gaussian process.
arXiv preprint arXiv:2108.10280
(2021)
\end{botherref}
\endbibitem

\bibitem{aglietti2020causal}
\begin{bchapter}
\bauthor{\bsnm{Aglietti}, \binits{V.}},
\bauthor{\bsnm{Lu}, \binits{X.}},
\bauthor{\bsnm{Paleyes}, \binits{A.}},
\bauthor{\bsnm{Gonz{\'a}lez}, \binits{J.}}:
\bctitle{Causal bayesian optimization}.
In: \bbtitle{International Conference on Artificial Intelligence and
  Statistics},
pp. \bfpage{3155}--\blpage{3164}
(\byear{2020}).
\bcomment{PMLR}
\end{bchapter}
\endbibitem

\bibitem{holler2014x}
\begin{barticle}
\bauthor{\bsnm{Holler}, \binits{M.}},
\bauthor{\bsnm{Diaz}, \binits{A.}},
\bauthor{\bsnm{Guizar-Sicairos}, \binits{M.}},
\bauthor{\bsnm{Karvinen}, \binits{P.}},
\bauthor{\bsnm{F{\"a}rm}, \binits{E.}},
\bauthor{\bsnm{H{\"a}rk{\"o}nen}, \binits{E.}},
\bauthor{\bsnm{Ritala}, \binits{M.}},
\bauthor{\bsnm{Menzel}, \binits{A.}},
\bauthor{\bsnm{Raabe}, \binits{J.}},
\bauthor{\bsnm{Bunk}, \binits{O.}}:
\batitle{X-ray ptychographic computed tomography at 16 nm isotropic 3d
  resolution}.
\bjtitle{Scientific reports}
\bvolume{4}(\bissue{1}),
\bfpage{1}--\blpage{5}
(\byear{2014})
\end{barticle}
\endbibitem

\bibitem{jeong2005efficient}
\begin{bchapter}
\bauthor{\bsnm{Jeong}, \binits{S.}},
\bauthor{\bsnm{Obayashi}, \binits{S.}}:
\bctitle{Efficient global optimization (ego) for multi-objective problem and
  data mining}.
In: \bbtitle{2005 IEEE Congress on Evolutionary Computation},
vol. \bseriesno{3},
pp. \bfpage{2138}--\blpage{2145}
(\byear{2005}).
\bcomment{IEEE}
\end{bchapter}
\endbibitem

\bibitem{clark2014dynamic}
\begin{barticle}
\bauthor{\bsnm{Clark}, \binits{J.N.}},
\bauthor{\bsnm{Huang}, \binits{X.}},
\bauthor{\bsnm{Harder}, \binits{R.J.}},
\bauthor{\bsnm{Robinson}, \binits{I.K.}}:
\batitle{Dynamic imaging using ptychography}.
\bjtitle{Physical review letters}
\bvolume{112}(\bissue{11}),
\bfpage{113901}
(\byear{2014})
\end{barticle}
\endbibitem

\bibitem{head2018scikit}
\begin{botherref}
\oauthor{\bsnm{Head}, \binits{T.}},
\oauthor{\bsnm{MechCoder}, \binits{G.L.}},
\oauthor{\bsnm{Shcherbatyi}, \binits{I.}}, et al.:
scikit-optimize/scikit-optimize: v0. 5.2.
Zenodo
(2018)
\end{botherref}
\endbibitem

\bibitem{genton2001classes}
\begin{barticle}
\bauthor{\bsnm{Genton}, \binits{M.G.}}:
\batitle{Classes of kernels for machine learning: a statistics perspective}.
\bjtitle{Journal of machine learning research}
\bvolume{2}(\bissue{Dec}),
\bfpage{299}--\blpage{312}
(\byear{2001})
\end{barticle}
\endbibitem

\bibitem{hoffman2011portfolio}
\begin{bchapter}
\bauthor{\bsnm{Hoffman}, \binits{M.}},
\bauthor{\bsnm{Brochu}, \binits{E.}},
\bauthor{\bparticle{de} \bsnm{Freitas}, \binits{N.}}, \betal:
\bctitle{Portfolio allocation for bayesian optimization.}
In: \bbtitle{UAI},
pp. \bfpage{327}--\blpage{336}
(\byear{2011}).
\bcomment{Citeseer}
\end{bchapter}
\endbibitem

\bibitem{kushner1964new}
\begin{botherref}
\oauthor{\bsnm{Kushner}, \binits{H.J.}}:
A new method of locating the maximum point of an arbitrary multipeak curve in
  the presence of noise
(1964)
\end{botherref}
\endbibitem

\bibitem{mockus1978application}
\begin{barticle}
\bauthor{\bsnm{Mockus}, \binits{J.}},
\bauthor{\bsnm{Tiesis}, \binits{V.}},
\bauthor{\bsnm{Zilinskas}, \binits{A.}}:
\batitle{The application of bayesian methods for seeking the extremum}.
\bjtitle{Towards global optimization}
\bvolume{2}(\bissue{117-129}),
\bfpage{2}
(\byear{1978})
\end{barticle}
\endbibitem

\bibitem{cox1992statistical}
\begin{bchapter}
\bauthor{\bsnm{Cox}, \binits{D.D.}},
\bauthor{\bsnm{John}, \binits{S.}}:
\bctitle{A statistical method for global optimization}.
In: \bbtitle{[Proceedings] 1992 IEEE International Conference on Systems, Man,
  and Cybernetics},
pp. \bfpage{1241}--\blpage{1246}
(\byear{1992}).
\bcomment{IEEE}
\end{bchapter}
\endbibitem

\bibitem{kirkland1998advanced}
\begin{bbook}
\bauthor{\bsnm{Kirkland}, \binits{E.J.}}:
\bbtitle{Advanced Computing in Electron Microscopy}
vol. \bseriesno{12}.
\bpublisher{Springer},
\blocation{New York}
(\byear{1998})
\end{bbook}
\endbibitem

\bibitem{hammersley1960monte}
\begin{barticle}
\bauthor{\bsnm{Hammersley}, \binits{J.M.}}:
\batitle{Monte carlo methods for solving multivariable problems}.
\bjtitle{Annals of the New York Academy of Sciences}
\bvolume{86}(\bissue{3}),
\bfpage{844}--\blpage{874}
(\byear{1960})
\end{barticle}
\endbibitem

\end{thebibliography}
\end{document}


\title{Supplementary Information}

\author[1]{Michael C. Cao}
\author[2]{Zhen Chen}
\author[3*]{Yi Jiang}
\author[1*]{Yimo Han}
\affil[1]{Department of Materials Science and NanoEngineering, Rice University, Houston, TX, USA 77005}
\affil[2]{School of Materials Science and Engineering, Tsinghua University, Beijing 100084, China}
\affil[3]{Advanced Photon Source, Argonne National Laboratory, Lemont, IL, USA 60439}

\maketitle
\section{Figures}
\begin{figure}[H]
    \centering
    \includegraphics[width = 0.75\linewidth]{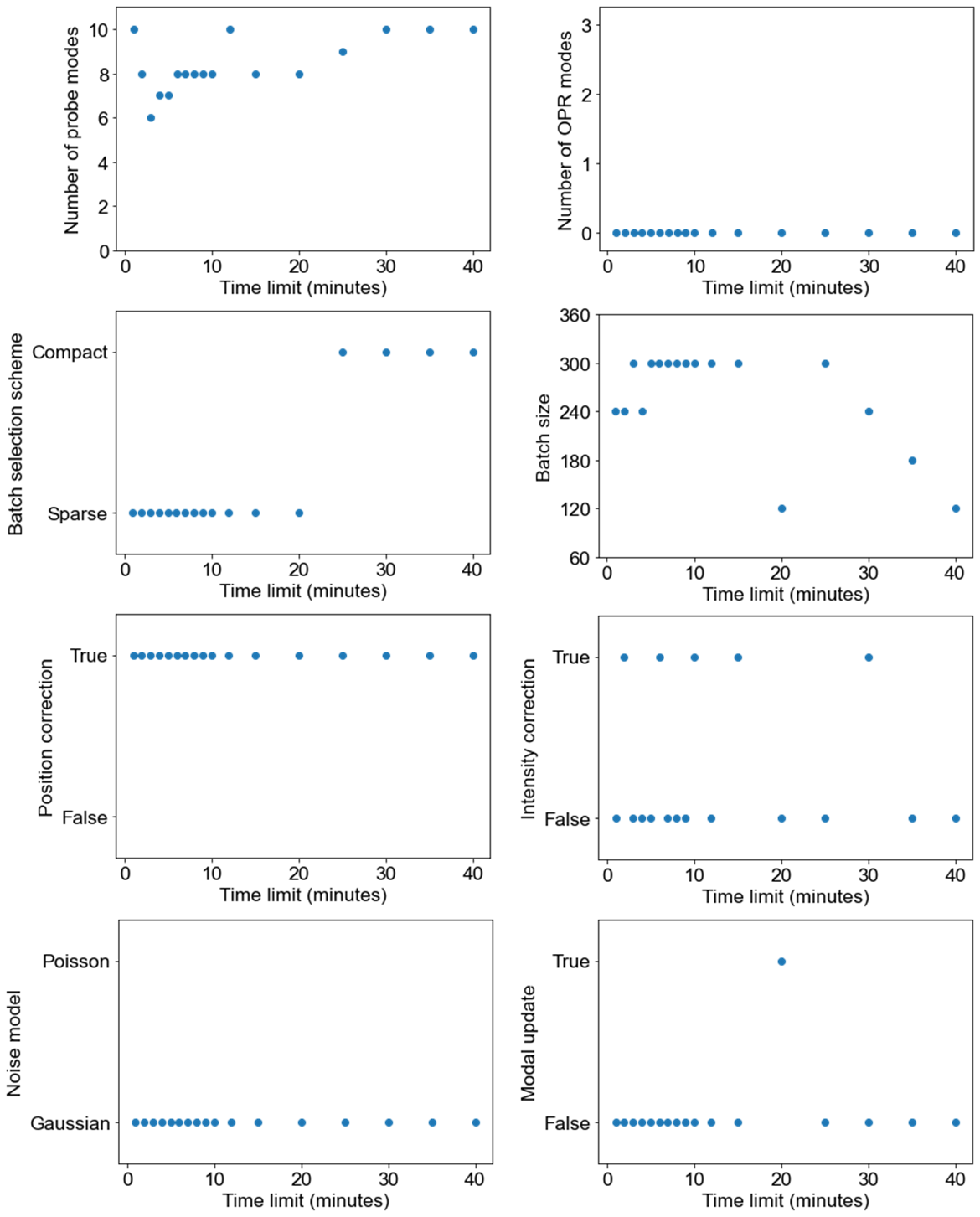}
    \caption{Optimal reconstruction parameters of an experimental dataset of bilayer MoSe$_2$/WS$_2$. At each time limit, eight different types of parameters were optimized by automatic parameter tuning with Bayesian optimization.}
    \label{fig:my_label}
\end{figure}

\begin{figure}[H]
    \centering
    \includegraphics[width = 0.75\linewidth]{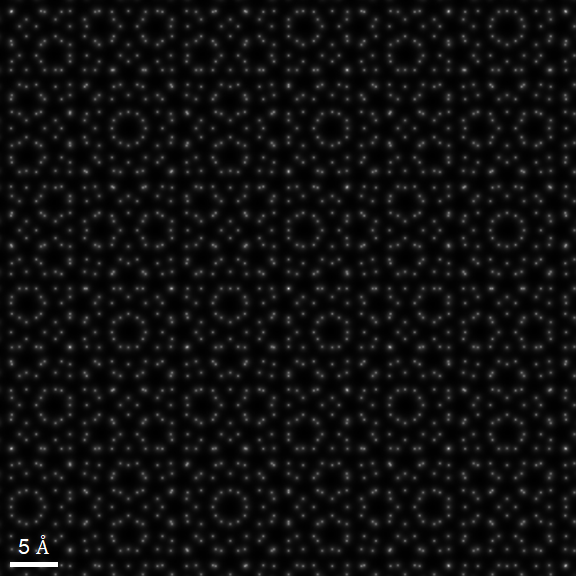}
    \caption{Projected potentials of simulated bilayer MoS$_2$ structure with a 30$\degree$ twist. The object was used to simulate electron ptychograpy data in automatic experimental parameter optimization.}
    \label{fig:my_label}
\end{figure}
\section{Tables}

\begin{table}[ht]
    \centering
    \begin{tabular}{|l|l|}
        \hline
        Reconstruction Parameter & Value \\
        \hline
        Number of mixed-state modes & 1, 2, 3, 4, 5, 6, 7, 8, 9, 10 \\
        \hline
        Batch size & 60, 120, 180, 240, 300 \\
        \hline
        Batch selection scheme & Sparse, Compact \\
        \hline
        Number of OPR modes & 0, 1, 2 \\
        \hline
        Intensity correction & True, False \\
        \hline
        Position correction & True, False \\
        \hline
        Modal update & True, False \\
        \hline
        Noise model & Gaussian, Poisson \\
        \hline
    \end{tabular}
    \caption{Reconstruction parameters and possible values in automatic parameter tuning. There are a total number of 4800 possible combinations.}
    \label{tab:recon_params}
\end{table}

\begin{table*}[ht]
    \centering
    \begin{tabular}{|l|l|l|l|l|}
         \hline
         Figure & 1-a,b & 1-c,d & 1-e,f & 1-g,h \\
         \hline
         \multicolumn{5}{|c|}{Reconstruction parameters}\\
         \hline
         Number of mixed-state probe modes & 7 & 7 & 7 & \underline{\textbf{1}} \\
         \hline
         Batch size & 300 & \underline{\textbf{60}} & 300 & 300 \\
         \hline
         Batch selection scheme & Sparse & Sparse & Sparse & Sparse \\
         \hline
         Number of OPR modes & 0 & 0 & 0 & 0 \\
         \hline
         Intensity correction & False & False & False & False \\
         \hline
         Position correction & True & True & \underline{\textbf{False}} & True \\
         \hline
         Noise model & Gaussian & Gaussian & Gaussian & Gaussian \\
         \hline
         Modal update & False & False & False & False \\
         \hline
         \multicolumn{5}{|c|}{Reconstruction quality}\\
         \hline
         Area under the FRC curve & 0.818 & 0.709 & 0.601 & 0.239 \\
         \hline
         1-bit FRC resolution (\AA) & 0.206 & 0.433 & 0.438 & 0.997 \\
         \hline
         SSIM & 0.896 & 0.816 & 0.791 & 0.518 \\
         \hline
    \end{tabular}
    \caption{Reconstruction parameters and quality evaluations of bilayer MoSe$_2$/WS$_2$ sample. The parameters correspond to the reconstructions that are shown in Figure 2. The automatic parameter tuning with BO was used to optimize the area under the FRC curve, which reflects the similarity between two independent reconstructions. This is consistent with visual inspection and other metrics such as the 1-bit FRC resolution and SSIM.}
    \label{tab:recon_param_quality}
\end{table*}

\begin{table}[ht]
    \centering
    \begin{tabular}{|l|l|}
        \hline
        Experimental parameter & Range \\
        \hline
        Scan step size (\AA) & 1.5 to 15 \\
        \hline
        Aperture size (mrad) & 5 to 35 \\
        \hline
        Probe defocus (nm) & 0 to 1000 \\
        \hline
        Detector size (\# of pixels) & 64x64, 128x128, 196x196, ..., 1024x1024 \\
        \hline
    \end{tabular}
    \caption{Experimental parameters and ranges in automatic parameter optimization for dose-limited electron ptychography. }
    \label{tab:my_label}
\end{table}